\documentclass[aps,preprint]{revtex4}
\usepackage{amssymb}
\usepackage{empheq}
\usepackage{exscale,pstricks}
\usepackage{wrapfig}

\begin{document}

\title{Effects of local heating and premelting in the terminal part of the e$^+$ track}

\author{ Dmitry S. Zvezhinskiy$^{a}$\footnote{zmitja@yandex.ru} and Sergey V. Stepanov$^{a}$ and
Vsevolod M. Byakov$^{a,b}$ and Bozena Zgardzinska$^{c}$}
\affiliation{$^{a}$ Institute of Theoretical and Experimental Physics, B.Cheremushkinskaya 25, 117218 Moscow, Russia \\
$^{b}$ D.Mendeleyev University of Chemical Technology, Miusskaya sq., 9, Moscow 125047, Russia \\
$^{c}$ Department of Nuclear Methods, Institute of Physics, Maria Curie-Sklodowska University, pl. M. Curie-Sklodowskiej 1, 20 031
Lublin, Poland}

\maketitle

\subsection{Introduction}
The terminal part of the e$^{+}$ track (the positron blob) is formed during ionization slowing down and subsequent
ion-electron recombinations produced by a positron. It releases up to 1 keV of energy, which is
converted into heat within few picoseconds. If a bulk temperature of a medium is below, but close enough to its
melting point, some region of a substance may melt, yielding a peculiar temperature dependence of the
lifetime (LT) spectra.

We have estimated properties of the molten region with a help of macroscopic heat conduction equation
and suggested a model describing temperature dependence of the ortho-positronium lifetime in frozen
methanol, ethanol, butanol and water close to their melting points.

\subsection{General equations. Homogeneous medium}
In a homogeneous medium the heat conductivity equation for the e$^+$
blob looks as follows:
\begin{equation}\label{LH1}
c_p \rho \frac{\partial T(r, t)}{\partial t} = \hbox{div}(\lambda \nabla T)  + q_+(r,t),
\qquad T(r,t=0) = T_{bulk}.
\end{equation}
Here $T(r,t)$ is the local temperature, $T_{bulk}$ is the bulk temperature of the medium, $c_p$ is its specific
heat capacity, $\rho$ is the density, $\lambda$ is the thermal conductivity. The second term in the RHS
describes energy release by the positron when it creates the blob: $q_+(r,t) \approx W_{blob} G(r,a)
f(t,\tau)$, where $W_{blob}\approx 1$ keV is the blob formation energy, $G(r,a) =
\frac{e^{-r^2/a^2}}{\pi^{3/2}a^3}$ describes spatial distribution of the released energy ($a$ is about tens of
\AA) and $f(t,\tau)$ its temporal distribution (here $\tau$ is the typical time of ion-electron
recombination and should be of the order of 1-10 ps). We adopted that $f(t,\tau) =
\exp\left(-\frac{(t-1~\text{ps})^2}{2 \tau^2}\right)/(\sqrt{2 \pi}\tau )$, where $\tau=0.3 \text{\,ps}$. Since
a system is simultaneously solid and liquid we used the following method for obtaining a numerical solution of
Eq. (\ref{LH1}) \cite{Muh09}. An additional contribution to $c_p(T)$ was used to simulate a presence of the
latent heat of melting ($q_m$). It is non-zero only in a small temperature
interval $\Delta T$ around $T_m$ (the melting point temperature) and its integral over temperature in
$\Delta T$ range should be equal to $q_m$. The value of $\Delta T$ (phase transition width) was arbitrarily fixed to
0.25 K.  An additional contribution to $c_p$ related with the latent heat was simulated by a Gaussian
$T$-dependent function.
Dependencies of $c_p(T)$ as well as $\lambda(T)$ and $\rho(T)$ are described by smooth
functions over $T$ as proposed in \cite{Muh09}.
Thermodynamical properties of methanol, ethanol, butanol and water used in present work are taken from \cite{cp_alcohols,Yaws,Kor11,dean}.

\subsection{Estimated properties of molten region}

Temperature profiles $T(r,t)$ were calculated numerically for $0<r<200$ \AA\ and $0<t<1$ ns. Eq. (\ref{LH1}) was solved as 1-D problem with boundary
conditions $T(r=200{\rm \AA}) =T_{bulk}$ with a help of \texttt{PDEPE} solver from \texttt{Matlab}. The
phase of medium is deduced from the temperature of each spatial point, which can be higher than $T_{m}+\Delta T$ (liquid phase) or lower than
$T_{m}-\Delta T$ (solid phase).

\begin{figure}[h]
 \centering
\vspace{-10pt}
\includegraphics[scale=0.5]{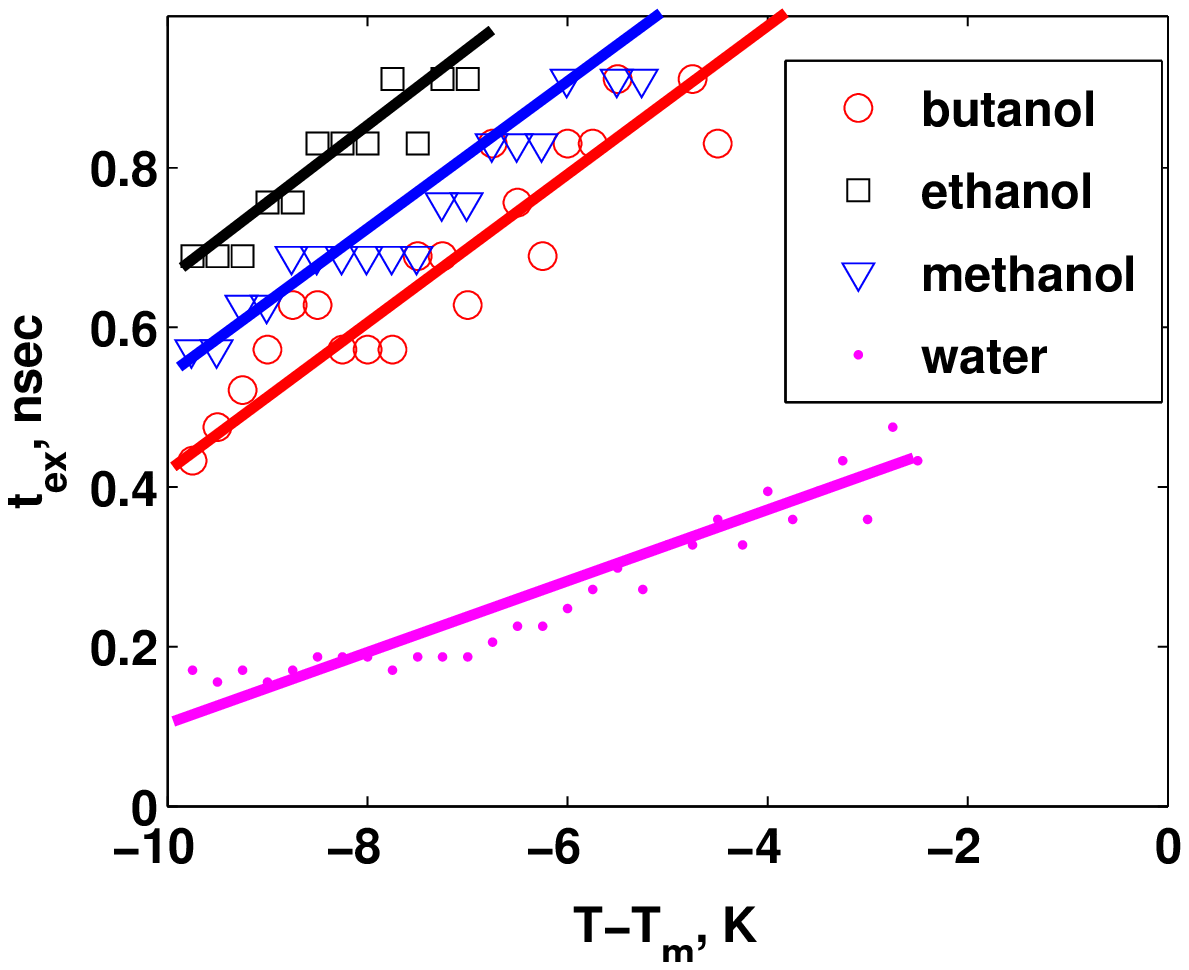}
 \includegraphics[scale=0.5]{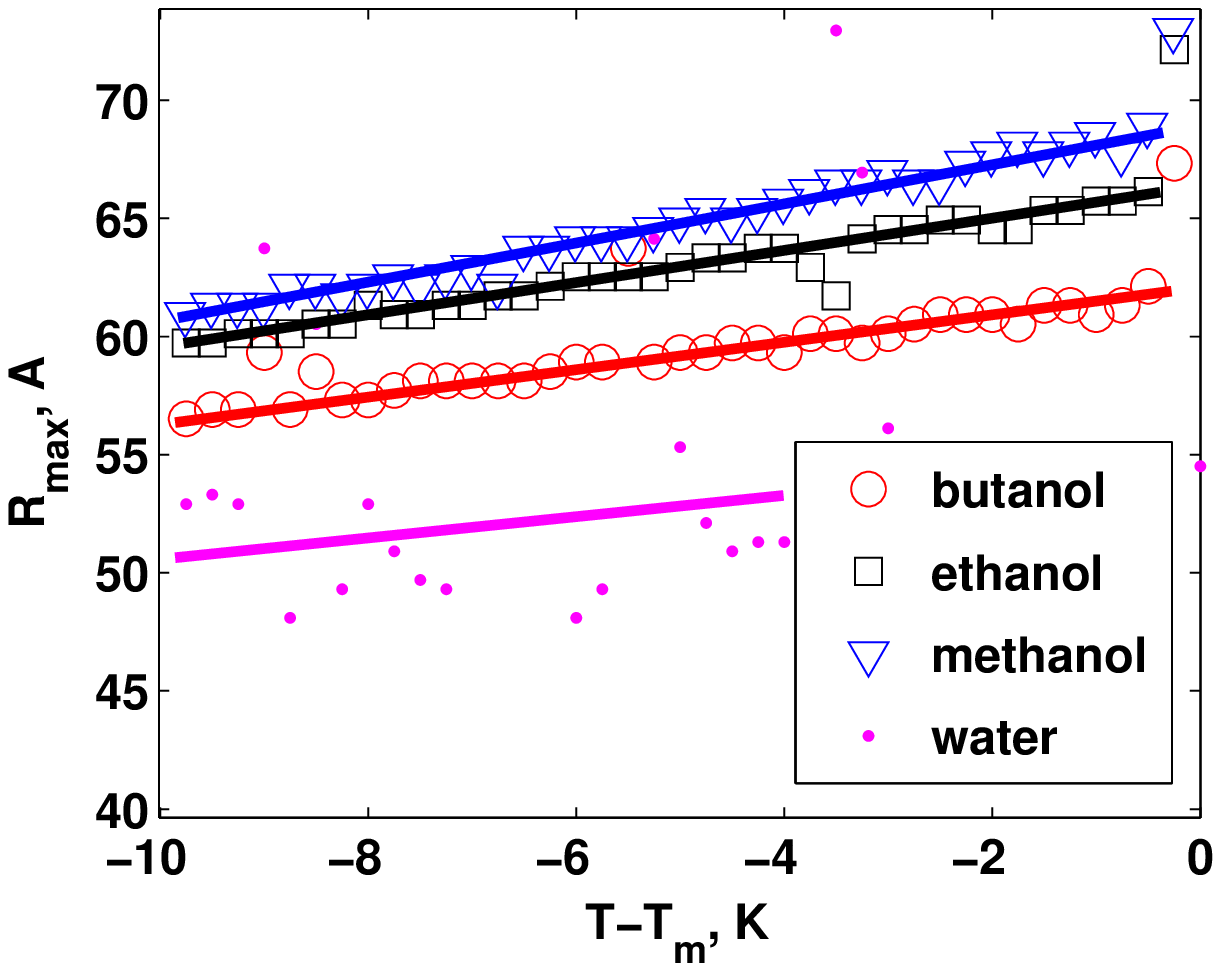}
\caption{Lifetime of the molten region $t_{m}$ (left) and its radius $R_m$ (right) for methanol,
etanol, buthanol and water are displayed as a function of $T_b-T_m$.
}
 \label{fig:features}
\vspace{-20pt}
\end{figure}

Fig. \ref{fig:features} displays the maximum radius of the molten region, $R_{max}$, vs. $T_{\rm bulk}$ and the lifetime $t_{max}$ of the molten region (at $t>t_{max}$ temperature of
an any point of the medium is below $T_m+\Delta T$).

\subsection{Experimental data.}

\begin{figure}{h}
 \centering
 \vspace{-10pt}
\includegraphics[scale=0.55]{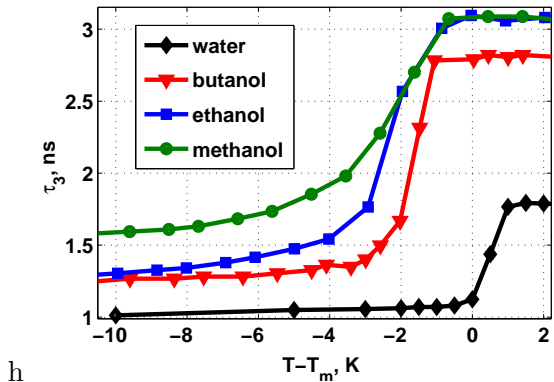}
 \caption{Experimental data for the lifetime of ortho-positronium component in alcohols and water.
 Temperature $T$ of investigated medium is plotted relatively to $T_m$ of certain substance.}
 \label{fig:T3exp}
\vspace{-20pt}
\end{figure}

Positron annihilation lifetime spectra in alcohols and in water were measured in 
Maria Curie-Sklodowska University. Each
lifetime spectrum was deconvoluted into 3 exponents by means of \texttt{LT 92} program. The data were processed at a fixed ratio
of ortho-Ps to para-Ps intensities equal to 3:1. Fig. \ref{fig:T3exp} displays temperature variation of the lifetime of the long-lived component of
LT spectra vs. $T$ close to the melting points.

For alcohols one can observe a noticeable temperature shift between the beginning of $\tau_3(T)$ plateau and
the melting point.
For description of $\tau_3(T)$ dependencies the following model is proposed.
\subsection{Theoretical model.}
Let us divide a space-time evolution of the Ps into two subsequent stages, namely, its formation and annihilation. The first stage is
initiated by a heat generation during e$^+$ ionization slowing down, when it forms a terminal blob. As a result, intrablob region may melt, so therein
formation of the Ps bubble is possible. Typical duration of the first stage may be estimated as about $t_{max}\sim 10$
ps, so annihilation may be neglected.

When temperature of the medium approaches to the melting point, concentration of free volume elements (FVE; or vacancy-like defects) rise extensively.
These FVE trap Ps and yield its typical lifetime as about $\tau_S \approx 1.2$-1.5 ps.

Frozen phase usually consists of domains which are disordered in a different extent (local density fluctuations,
intercrystalline boundaries and so on). Local inhomogeneity of the medium leads to smearing of its melting temperature. It is the reason of the
{\it premelting} effect (an appearance of some molten domains when $T_{bulk}$ is slightly below $T_m$). Hence the formation of the Ps bubble state may occur in these premelted regions.

Let us consider Ps formation during the first stage in terms of concentrations of different Ps states,
namely, mobile quasi-free positronium qf-Ps, $c_{qf}(r,t)$, Ps trapped in FVE of a solid phase, $c_v$, and Ps bubble state, $c_b$, formed
in the molten region of the blob and in the premelted domains. Evolution of these states is described by the following set of kinetic equations:
\begin{equation}
\frac{\partial c_{qf}(r,t)}{\partial t} = D \Delta c_{qf} - (\lambda_v + \lambda_b) c_{qf},
 \qquad
\frac{\partial c_v(r,t)}{\partial t} = \lambda_v c_{qf},
 \qquad
\frac{\partial c_b(r,t)}{\partial t} = \lambda_b c_{qf}.
 \label{eq:pd1}
\end{equation}
Here $D \sim 0.03$ cm$^2$/sec is the diffusion coefficient of qf-Ps (we assume that the mean free path of Ps transport is about its de Broglie wave length
and qf-Ps velocity is the thermal one), $\lambda_v(T)$ and $\lambda_b(T)$ are Ps trapping rates by FVE and premelted regions.

At $t=0$ we have qf-Ps only. We adopt that its initial distribution is approximately Gaussuan and that it remains Gaussian later at any $t$,
namely, $c_{qf}(r,t) = n_{qf}(t) G(r,t,D)$, where $G(r,t,D)=\frac{\exp{\left( -r^2/(a^2+4Dt) \right)}}{\left[ \pi (a^2+4Dt) \right] ^{3/2}}$. Hence $n_{qf}(t) = P_{\rm qf-Ps}e^{-(\lambda_v +\lambda_b) t}$, where $P_{\rm qf-Ps}$ is the formation probability of qf-Ps. Distributions $c_v$ and $c_b$ can be found from (\ref{eq:pd1}) and are equal to $c_{v,b}(r,t) = \int_{0}^{t} dt'\, \lambda_{v,b}\,
c_{qf}(r,t')$.

The first stage is terminated by a formation of the molten region with the radius, attaining its maximal value $R_m(T_{bulk})$ at
$t_{max}$. By this time the fraction $f_m$ of Ps atoms in premelted regions (i.e. in a liquid phase) is
$ f_m = \int_{r<R_m} (c_{qf}(r,t_{max}) +c_v(r,t_{max}) +c_b(r,t_{max}))d^3r$. By the time $t_{max}\sim 10$ ps all these Ps states turn out to be in
a bubble state (Ps bubble formation time is about 10 ps as well \cite{Mik07}). After subsequent freezing of the molten region we assume that the bubble
state does not collapse. Cavity survives after freezing with the Ps inside, so ortho-Ps lifetime remains the same.

We assume that all qf-Ps atoms outside the molten region, remaining quasi-free by the time $t_{max}$, become trapped shortly by FVE or
premelted regions ($n_v$ or $n_b$) in proportion to $\lambda_v/(\lambda_v+\lambda_b)$ and $\lambda_b/(\lambda_v+\lambda_b)$. Therefore, total formation
probability $P_{\rm Ps}$ of the Ps bubble state arises not only from the molten central region of the blob, but also
from premelted domains, so that $P_{\rm Ps} = f_m + \int_{r>R_m} \left[ c_b(r,t_{max}) + \frac{\lambda_b}{\lambda_v +\lambda_b} c_{qf}(r,t_{max})
\right] d^3r$.

Annihilation of the free positrons and Ps states occurs on the second stage at $t>t_{max}\approx 0$. We assume that formation of para- and ortho-Ps spin
states takes place in a conventional 1:3 proportion (possible influence of intratrack chemical reactions which may change this ratio is neglected). Let
$n_{\rm pPs~(oPs)}$ be the fraction of para-Ps (ortho-Ps) formed in a bubble state.
Following equations allow to calculate a shape of LT spectrum according to proposed model:
$$
\dot n_+ = -\lambda_+ n_+, \qquad n_+(0) = 1-P_{\hbox{\scriptsize qf-Ps}};
 \qquad
\dot n_{\rm oPs} = -\lambda_{po} n_{\rm oPs}, \qquad n_{\rm oPs}(0) = 3P_{\rm Ps}/4;
$$
$$
\dot n_{\rm pPs} = -\lambda_{\rm pPs} n_{\rm pPs}, \qquad
 n_{\rm pPs}(0) = \frac{P_{\rm Ps}}{4} + \frac{1}{4} \int_{r>R_m} \left( c_v(r,t_{max}) + \frac{\lambda_v}{\lambda_v+\lambda_b} c_{qf}(r,t_{max})\right) d^3r;
$$
$$
\dot n_v = -\lambda_S n_v, \qquad
n_v(0) = \frac{3}{4} \int_{r>R_m} \left(c_v(r,t_{max})+\frac{\lambda_v}{\lambda_v+\lambda_b} c_{qf}(r,t_{max})\right) d^3r.
$$

These equations yield a four-component LT spectrum, where $\lambda_+ \approx 2$ ns$^{-1}$ is the free e$^+$ annihilation rate, $\lambda_{\rm pPs}$ is the
annihilation rate of para-Ps (we do not distinguish p-Ps either in a bubble state or not), $\lambda_{po}$ is the pick-off annihilation rate of Ps in a
bubble state and $\lambda_S$ is the o-Ps annihilation rate in free volume elements of frozen regions.
However, by now experimental data were deconvoluted into three exponents. So we used the following expression to
describe the long-lived component $\tau_3$ of the spectra:
\begin{equation}\label{eq:T3}
 \frac{1}{\tau_3} = \lambda_S \frac{n_v(0)}{n_v(0)+n_{\rm oPs}(0)} +
					\lambda_{po} \frac{n_{\rm oPs}(0)}{n_v(0)+n_{\rm oPs}(0)}.
\end{equation}
Implicitly two temperature dependent quantities, $\lambda_v$ and $\lambda_b$, enter this expression. It is
reasonable to describe them by the Arrhenius-like law $\lambda_{v,b}(T) \sim \exp (-E_{v,b}/T)$ and extract
corresponding activation energies for free volume elements $E_v$ and premelted regions $E_b$.

\begin{figure}{h}
 \centering
 \includegraphics[scale=0.55]{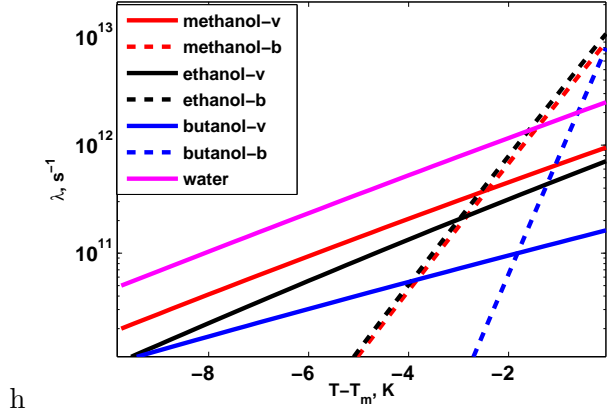}
 \caption{Temperature dependencies of trapping rates $\lambda_v$ (solid lines) and $\lambda_b$ (dashed lines) for methanol, ethanol, butanol and water. Temperature is given with respect to melting point of each substance.}\label{fig:trrate}
\vspace{-15pt}
\end{figure}

Fitting the model (\ref{eq:T3}) to the experimental data (Fig. \ref{fig:T3exp})
 in the range close to melting points, we obtain temperature dependencies of trapping rates,
Fig. \ref{fig:trrate}. To eliminate correlation of parameters in the Arrhenius law the following procedure was applied. Melting is assumed to occur
when the number of premelted regions (let a typical size of these regions is 
$R_S$) allows to fill the whole space, i.e. $\lambda_b(T=T_m)/\left[4\pi D R_S\right]\cdotp\frac{4}{3}\pi R_S^3=1$. From this relation it is possible to get an additional constraint for $\lambda_b(T=T_m)$. Furthermore, if
we consider the energy $E_b$ as an energy necessary for melting of the media inside a premelted domain ($\frac{4}{3}\pi R_S^3$), then $E_b=q_m \cdotp
\frac{4}{3}\pi R_S^3 \cdotp \rho_S$. Thus obtained relations for $E_b$ and $\lambda_b(T=T_m)$ are dependent on the only parameter $R_S$ which has a meaning of
geometrical size of homogeneously distributed volumes responsible for melting.

As a result of fitting we obtain activation energies which are as follows:
\begin{center}
\begin{tabular}{c|c|c|c|c}
substance & methanol & ethanol & butanol & water\\
$E_v$, [eV] & 1 & 0.9 & 0.8 & 2.5 \\
$E_b$, [eV] & 3.5 & 2.9 & 7.2 & Not used
\end{tabular}
\end{center}

According to experimental data (Fig. \ref{fig:T3exp}), water doesn't indicate any significant influence of
premelting on positronium lifetime especially before its melting point, therefore for it we assumed $\lambda_b=0$.

This work is supported by the Russian Foundation of Basic Research (grant 11-03-01066) and Federal Agency on Atomic Energy.

\end{document}